\documentclass[12pt,preprint]{aastex}

\shorttitle{EUV Non-thermal Line Broadening and High-energy particles during Solar Flares}
\shortauthors{Kawate \& Imada}

\begin{document}

\title{ The Relationship Between EUV Nonthermal Line Broadening and High-energy Particle During Solar Flare}

\author{T. \textsc{Kawate},\altaffilmark{1} \altaffilmark{2}
S. \textsc{Imada},\altaffilmark{3} \altaffilmark{2}
}
\altaffiltext{1}{Kwasan and Hida Observatory, Kyoto University, Kurabashira, Kamitakaracho, Takayama, Gifu 506-1314,Japan}
\altaffiltext{2}{ National Astronomical Observatory of Japan, 2-21-1 Osawa, Mitaka, Tokyo 181-0015, Japan}
\altaffiltext{3}{ Solar-Terrestrial Environment Laboratory, Nagoya University, Furocho, Chikusa-ku, Nagoya, Aichi 464-8601, Japan}

\begin{abstract}
We have studied the relationship between the location of EUV nonthermal broadening and high-energy particles during the large flares by using EUV imaging spectrometer onboard {\it Hinode},  Nobeyama Radio Polarimeter, Nobeyama Radioheliograph, and Atmospheric Imaging Assembly onboard {\it Solar Dynamic Observatory}.
We have analyzed the five large flare events which contain thermal rich, 
intermediate,
and thermal poor flares classified by the definition discussed in the paper. 
We found that, in the case of thermal rich flares, the nonthermal broadening of \ion{Fe}{24} occurred at the top of the flaring loop at the beginning of the flares.
The source of the 17 GHz microwave is located at the footpoint of the flare loop.
On the other hand, in the case of intermediate/thermal poor flares, the nonthermal broadening of \ion{Fe}{24} occurred at the footpoint of the flare loop at the beginning of the flares. The source of the 17 GHz microwave is located at the top of the flaring loop.
We discussed the difference between thermal rich and intermediate/thermal poor flare based on the spatial information of nonthermal broadening, which may give a clue for the presence of turbulence playing an important role in the pitch angle scattering of the high-energy electron.

\end{abstract}


\section{Introduction}
Particle acceleration has been discussed as one of the long-standing problem in astrophysical plasma.
It is often observed that the high-energy particles are generated with the explosive energy release such as the supernova explosion, the solar flare, and the substorm in the Earth's magnetosphere.
Several observations indicate that the energy spectrum in the high energy range can be described by the power law distribution.
Many theoretical and observational studies have been carried out to understand the origin of high-energy particles, and various mechanisms have been proposed, such as the shock acceleration \citep[e.g.,][]{bla}.
Recently, magnetic reconnection is thought to be important for particle acceleration, because high-energy particles can be observed well in association with magnetic reconnection.
For example, in the Earth's magnetosphere the energetic particles are often observed in the vicinity of magnetic reconnection region \citep[e.g.][]{sar,oie,ima,ima2}.

Solar flares, which is also believed to be associated with magnetic reconnection \citep[e.g.,][]{car,stu,hir,kop,pne},  is another important field for particle acceleration.
The solar atmosphere is an excellent space laboratory for particle acceleration because of the observability on a large scale.
Over the last several decades, considerable effort has been devoted toward understanding high-energy particles during a solar flare. 
In the last two decades there have been substantial advances in the study of particle acceleration, mainly due to modern satellite observation in hard X-ray (HXR), 
for example, the Hard X-ray Telescope oboard Yohkoh \citep{kosugi1991} and RHESSI \citep{lin2002}.  
\cite{mas} discussed the location of HXR sources, and they concluded that the HXR sources are located above the flare loops which is observed in the soft X-ray (SXR) range.
This result indicates that the magnetic reconnection occurred above the SXR flare loops, and the high-energy particles might be accelerated at the downstream of the reconnection outflow.
After \cite{mas}, several discussions of the particle acceleration mechanisms have been intensively carried out.
There are essentially three major mechanism for particle acceleration during solar flares: 1) DC electric field acceleration \citep[e.g.,][]{kli}, 2) stochastic acceleration \citep[e.g.,][]{mel}, and 3) shock acceleration \citep[e.g.,][]{tsu}. 
The most plausible mechanism to explain the observed high-energy particles is still under discussion.
However, in most mechanism, even including those based on DC electric field acceleration \citep[e.g.,][]{amb}, the wave and/or turbulence have an important role for particle acceleration.
Therefore, many people believed that understanding  the characteristics of wave and/or turbulence in the course of solar flares is crucial for understanding the particle acceleration mechanism.

The wave and/or turbulence in the solar corona are often discussed from the spectroscopic observation in the EUV or SXR wavelength range. 
The spectral line width obtained by the spectroscopic observation mainly consists of three parts: 1) thermal width, 2) instrumental width, 3) nonthermal width.
It is generally believed that the nonthermal width is associated  with wave/turbulence or velocity gradients \citep[e.g.,][]{dos,ima3}.
Nonthermal width is expressed by a following formula,
\begin{eqnarray}\label{e1}
W_{N}=\sqrt{W_{obs}^2-W_I^2-4\ln 2 \frac{2kT_i}{M_i}},
\end{eqnarray} 
where $W_{N}$ is nonthermal width, $W_{obs}$ is observed line width, $W_I$ is instrumental line width, 
$k$ is Boltzmann constant, $T_i$ is ion temperature, and $M_i$ is ion mass.
Nonthermal line broadening during solar flares in SXR emission line ($\sim 10^7$ K) has been reported by {\it Solar Maximum Mission} (SMM)/Bent Crystal Spectrometer (BCS) (\cite{rapley1977}) or {\it Yohkoh}/Bragg Crystal Spectrometer (BCS) (\cite{culhane1991}) observations. 
After its finding, the characteristics of the nonthermal line broadening have been intensively studied to clarify its origin.
In \cite{mariska1993}, they divided the flares into rising and decaying phase for 219 flare events observed by {\it Yohkoh}/BCS, and they concluded that there is no correlation between the line width of  \ion{Ca}{19} and the heliocentric distance, the peak intensity, the rising phase duration nor the total flare duration. 
On the other hand, there is a weak correlation between the line width and the Doppler-shift at the rising phase of the flares.
\cite{mariska1999} studied the 48 limb flares which contain occulted and non-occulted flares, and they found that there is no difference in the line width between the occulted and non-occulted flares.
\cite{harra1997} analyzed the small GOES class flares and found that the line width correlates with neither the flare size and the complexity, nor the HXR intensity.
 However, the line width weakly correlates with the total and rising duration of the flares.
From observational results of {\it SMM}/BCS and {\it Yohkoh}/BCS, \cite{ranns2000} proposed that the origin of nonthermal line broadening during solar flares is the turbulence at somewhere of the following positions: 
1) reconnection site, 2) above-the-looptop (HXR source), 3) SXR looptop, 4) evaporating chromospheric plasma, 
5) flare loop footpoint.
Further, \cite{ranns2001} analyzed 59 limb flares by using \ion{Ca}{19} line of {\it Yohkoh}/BCS and HXR of Burst And Transient Source Experiment (BATSE).
They found that the line broadening occurs after the HXR burst in the case of impulsive flares.
On the other hand, the line broadening occurs before the HXR burst in the long duration flares. 
Thus they concluded that the line broadening comes from turbulent evaporation at the footpoint in the case of impulsive flares.
However, the line broadening comes from looptop turbulence during the preflare phase in the case of long duration flare.

So far, various observational studies of nonthermal line broadening have been reported. 
However, the most of studies are not based on spatially resolved observations but integrated observation over the sun. 
Thus, there are no observational studies which directly discuss the relationship between the location of the EUV nonthermal line broadening and high-energy particles during solar flares. Recently, the Hinode spacecraft  \citep[][]{kosugi2007} onboard three telescopes, the Solar Optical Telescope (SOT), the X-Ray Telescope (XRT), and the EUV Imaging Spectrometer (EIS), was launched. EIS onboard {\it Hinode} is a high spectral resolution spectrometer aimed at studying dynamic phenomena in the corona with high spatial resolution and sensitivity  (\cite{culhane2007}). 
Thus, by using EIS observation we can directly discuss the location of the nonthermal line broadening.
\cite{watanabe2008} found the line broadening at the looptop during the C4.2 flare with the imaging spectroscopy observation by EIS. 
\cite{hara2008} also reported the line broadening of \ion{Ca}{17} at the looptop in the C class long durational event. 
On the other hand, \cite{imada2008} and \cite{milligan2011} found the line broadening at the footpoint. 
They argued that the origin of the line broadening might be the chromospheric evaporation.

In this paper, we will report the relationship between the location of EUV nonthermal broadening and high-energy particles during the large flares.
We also pay attention to the relationship between the location of nonthermal broadening and the flare characteristics such as the thermal richness and/or the duration of flares.
In section 2, we introduce the flare data and the result of analyses in the two typical flares.
In section 3, we show the results of other flare events to ensure the results which is discussed in section 2, and discuss the relationship of the location between nonthermal broadenng and high-energy electrons.
In section 4, we summarize our observational result and discuss the interpretation of the relationship between the nonthermal broadening and high-energy particles.

\section{Observation}
\subsection{Data}

We carry out the comprehensive analysis to clarify the relationship between the location of EUV nonthermal broadening and high-energy particles during the large flares.
We analyze the multi-wavelength data obtained by using {\it Hinode}/EIS,  Nobeyama Radioheliograph (NoRH; \cite{nakajima1994}), Nobeyama Radio Polarimeter (NoRP;  \cite{nakajima1985}, \cite{torii1979}), and {\it Solar Dynamics Observatory} (SDO)/Atmospheric Imaging Assembly (AIA) (\cite{lemen2012}).
By using Hinode Flare Catalogue\footnote{http://st4a.stelab.nagoya-u.ac.jp/hinode\_flare/} \citep[][]{wat}, we search the large solar flare event (larger than GOES M class) observed simultaneously by the instruments raised above from 2010 to 2012. 
We carefully check the data from the onset phase to the impulsive phase and exclude the event which maximum brightness temperature of 17GHz is less than $10^8$ K because of the failure level of image reconstruction. We also use Nobeyama Radioheliograph 
Event List\footnote{http://solar.nro.nao.ac.jp/norh/html/event/} to avoid very low flux event at microwave frequencies.
In the result, we find five events which is suitable for our purpose.
We label the events as (a), (b), (c), (d), (e) from the order of the time (Table 1). 
Figure 1 shows the lightcurves of GOES 1.0 - 8.0 \AA\  SXR flux and NoRP 17 GHz flux during the flare events.
It can be seen that the events (a), (b), (e) are impulsive flares which has a sharp peak at microwave frequencies, but the events (c), (d) are long duration flares.

All EIS data which we use in this paper are the sparse raster scanning observations.
The scanning step is 5 arcsec, although the slit width is 2 arcsec.
The exposure time is 8 sec except for the one flare event (2012 July 14th: 9 sec). 
It takes $\sim$ 6 minutes to obtain one scanning images (180$\times$152 arcsec$^2$).
EIS data from the raster are processed using the EIS team provided software ({\it eis\_prep}), which corrects for the flat field, dark current, cosmic rays, and hot pixels.
The slit tilt was corrected by {\it eis\_tilt} correction.
Due to the thermal deformation of the instrument, there is an orbital variation of the line position causing an artificial Doppler shift of +/- 20 km s$^{-1}$ which follows a sinusoidal behavior. 
This orbital variation of the line position was corrected by using the house keeping data \citep[][]{kam}.
We assume the instrumental line width ($W_I$) in Equation 1 is equal to 0.061 \AA\ for the short wavelength band and 0.062 \AA\ for the long wavelength band.
This assumption is not sensitive to our results, because the nonthermal line width in the high temperature range is large enough to be distinguished from the instrumental width during solar flares. 
In this paper we only use the \ion{Fe}{24} line of 192.03 \AA\ for EIS data, because we focus on the characteristics of nonthermal broadening in the high temperature. 
\ion{Fe}{24} is one of the highest temperature line observed by EIS.
The typical temperature of \ion{Fe}{24} is 2$\times$10$^7$ K.
Therefore, we can compare our results with the past results which discuss the nonthermal broadening by using the high temperature emission lines \citep[e.g., ][\ion{Ca}{19}: $T_e$ $\sim$ 3$\times$10$^7$ K)]{ranns2001}.

We use AIA 131 \AA\ data to discuss the temporal evolution of the flare loop.
The 131 \AA\ images contain \ion{Fe}{8} and \ion{Fe}{21}
lines, and \ion{Fe}{21} emissions generally dominates during the large flares. 
Therefore we can easily compare the AIA 131 \AA\ and EIS 192.03 \AA\ images, because the temperature sensitivities are almost the same.
We also use AIA 1700 \AA\ data to trace the flare ribbons.
All AIA data are processed by {\it aia\_prep}.

The characteristics of high-energy particles can be discussed by using NoRH and NoRP data.
NoRH data is taken in every 1 sec with the spatial resolution of about 10 arcsec.
The 17 GHz microwave flux are emitted from high-energy electrons trapped within the flare loop through the gyrosynchrotron radiation process.
Thus we can discuss the temporal evolution of high-energy particles from NoRH 17 GHz flux.
There are no spatial resolution for NoRP data.
Although the temporal resolution of the data is 0.1 sec, we integrate the data for 10 sec to reduce the noises.

We use the GOES SXR flux and the NoRH 17 GHz microwave data to classify the thermal richness of the flare event. 
From \cite{kawate2011} there is a positive correlation between SXR flux and 17 GHz flux in the logarithmic scale, and they discussed the thermal richness from the ratio between the GOES SXR flux and the NoRH 17 GHz microwave.
\cite{kawate2011} defined the Thermal Emission Index (TEI) to discuss the thermal richness of flares as the following expression, 
\begin{eqnarray}
TEI = \log _{10} \frac{I_{\rm GOES}}{{I_{\rm 17GHz}}^{0.59} } + 6.16.
\end{eqnarray} 
In the statistical study of \cite{kawate2011}, generally long duration events tend to be positive TEI.
On the other hand, short duration events tend to be negative TEI. 
In this paper, we use TEI for classification of the flare thermal richness.
We define 
TEI $>$ 0.2 as thermal rich, 0.2 $>$ TEI $>$ -0.2 as intermediate, and -0.2 $>$ TEI as thermal poor flares.
The flare events (a) and (b) can be classified as intermediate flares, the events (c) and (d) as thermal rich flares, and the event (e) as thermal poor flares (Table 1).
 
\subsection{ Intermediate Flare (event b)}
On 23 September 2011, the large solar flare (GOES M1.9, peak time 23:56) occurred at the northeast part of the sun (12$^\circ$ N, 56$^\circ$ E).
The maximum NoRH 17GHz flux was 225 sfu.
Lightcurves of NoRP 17 GHz, GOES 1.0 \AA\ - 8.0 \AA, and {\it RHESSI} 25 - 50 keV are shown in Figure 2 with solid line, dashed line, and dotted line, respectively.
The lightcurve of GOES has a sharp rising phase ($\sim$ 3 minutes), which is the typical characteristics of impulsive flare event.
The lightcurve at microwave frequencies or HXR has a sharp peak ($\sim$ 1 minutes), which is also the typical characteristics of impulsive flare event.
The TEI estimated by Equation 2 is 0.05 which can be classified as intermediate flare event.

First, we analyze the temporal evolution of high temperature plasma in the entire flare region by using \ion{Fe}{24} 192.03 \AA .
We integrate the line profile over the EIS field of view (FOV).
Figure 3 shows  an example of the line profile integrated over EIS FOV with Gaussian fitting during the flare.
We can obtain the intensity and nonthermal velocity from the integrated line profile.
We call them as full-FOV intensity and full-FOV nonthermal velocity.  
The lineprofile is integrated from 191.90 to 192.15 \AA\ to obtain the intensity.  
The full-FOV intensity and full-FOV nonthermal velocity is suitable for describing the temporal evolution of the entire flare plasma, although they do not have any spatial information.
Further, they can easily compare with the past results obtained from not spatially resolved telescope such as {\it Yohkoh}/BCS. 
Figure 4 shows the temporal variation of the full-FOV intensity (solid line), the full-FOV nonthermal velocity (horizontal line), and the microwave (dotted line). 
The length of the horizontal line in Figure 4 (full-FOV nonthermal velocity) represents the scanning duration for one raster image. 
The nonthermal velocities are plotted when the line peak is over 1000 erg cm$^{-2}$ sec$^{-1}$ str$^{-1}$ \AA$^{-1}$ to avoid the overestimation of line width by the contamination of \ion{Fe}{8} and \ion{Fe}{11} around 192.03 \AA.
As we mentioned before, \ion{Fe}{24} 192.03 \AA\ line is very strong during the flare.
Therefore, with the criteria we can easily avoid the contamination effect.
The full-FOV nonthermal velocity of \ion{Fe}{24} shows larger than 200 km sec$^{-1}$ during the microwave burst, and it gradually reduced with time.
On the other hand, the full-FOV intensity of \ion{Fe}{24} peaked after the microwave burst.
The light curve of GOES also peaked after the microwave burst (23:56). 
We cannot decide whether the full-FOV nonthermal velocity peaked before or  after the microwave burst, because the EIS raster scanning took longer time than the typical microwave burst duration.
However, it is remarkable that the full-FOV nonthermal velocity peaked before the peak of the full-FOV intensity. 

Next, we study the \ion{Fe}{24} line profiles which are not spatially integrated.
We are interested in the location of the nonthermal broadening at this time.
Figure 5 shows the intensity, the velocity, and the nonthermal velocity of \ion{Fe}{24} calculated by the Gaussian fitting during the impulsive phase.
We calculate the nonthermal velocity only in the case that the intensity is larger than 5$\times$10$^5$ erg cm$^{-2}$ sec$^{-1}$ str$^{-1}$ \AA$^{-1}$ 
to avoid the contamination of \ion{Fe}{8} and \ion{Fe}{11}, which we discussed above already.
In Figure 5, green contours show the nonthermal velocity of 100, 140, 180 km sec$^{-1}$. 
The maximum intensity and the maximum nonthermal velocity are almost cospatial.
Further, the northern part of the nonthermal broadening is roughly cospatial with the blueshift region ($\sim$ 150 km sec$^{-1}$) in the velocity of Figure 5 . 

Finally, we study the relationship of the location among the nonthermal broadening, the flare loop, and the flare ribbon.
We also discuss the temporal evolution of the relationship.
The temporal evolution of the relationship among the flare loop, the flare ribbon, and the nonthermal broadening is shown in Figure 6.
The color image, the green contour, and the black contour show AIA 131 \AA\ intensity, the nonthermal velocity of EIS 192.03 \AA, and AIA 1700 \AA\ intensity, respectively. 
The contour level of the nonthermal velocity is flexible.
In each panel of Figure 6(i-iv), the EIS raster scanning starts from 23:49:33, 23:55:23, 00:01:12, and 00:07:01, respectvely.
At the begining of the flare ( $\sim$ microwave peak), the nonthermal broadening (green contour) and the flare ribbons (black contour) were cospatial.
The location of nonthermal broadening gradually moved  toward the looptop.  

\subsection{ Thermal Rich Flare (event d)}

On 25 September 2011, the large solar flare (GOES M7.4, peak time 04:50) occurred at the northeast part of the sun (13$^\circ$ N, 50$^\circ$ E).
The maximum NoRH 17GHz flux was 283 sfu.
Lightcurves of NoRP 17 GHz, GOES 1.0 \AA\ - 8.0 \AA, and {\it RHESSI} 25 - 50 keV are shown in Figure 7 with solid line, dashed line, and dotted line, respectively.
The lightcurve of GOES has a gradual rising phase ($\sim$ 20 minutes), which is the typical characteristics of long duration flare event.
The lightcurve at microwave frequencies or HXR has a gradual peak ($\sim$ 5 minutes), which is also the typical characteristics of long duration flare event.
The TEI estimated by Equation 2 is 0.58 which can be classified as thermal rich flare event.

We analyze the thermal rich flare in the same way as the intermediate flare discussed in the previous subsection. 
Figure 8 shows  an example of the line profile integrated over EIS FOV with Gaussian fitting during the flare.
Figure 9 shows the temporal variation of the full-FOV intensity (solid line), the full-FOV nonthermal velocity (horizontal line), and the microwave (dotted line). 
The full-FOV nonthermal velocity of \ion{Fe}{24} gradually increased during the rising phase of the flare and reached to $\sim$150 km sec$^{-1}$ during the microwave burst.
After the microwave burst, it gradually reduced with time.
The full-FOV intensity of \ion{Fe}{24} (04:46) and GOES (04:50) also peaked at the microwave burst.
We cannot decide whether the full-FOV nonthermal velocity peaked before or  after the microwave burst, because the EIS raster scanning took longer time than the typical microwave burst duration.
However, it is remarkable that the full-FOV nonthermal velocity and the full-FOV intensity of \ion{Fe}{24} simultaneously peaked around the microwave burst. 

Next, we study the \ion{Fe}{24} line profiles which are not spatially integrated.
Figure 10 shows the intensity, the velocity, and the nonthermal velocity of \ion{Fe}{24} calculated by the Gaussian fitting during the impulsive phase.
We calculate the nonthermal velocity only in the case that the intensity is larger than 5$\times$10$^5$ erg cm$^{-2}$ sec$^{-1}$ str$^{-1}$ \AA$^{-1}$ to avoid the contamination of \ion{Fe}{8} and \ion{Fe}{11}, which we discussed above already.
In Figure 10, green contour shows the nonthermal velocity of 100, 140, 180 km sec$^{-1}$. 
The maximum nonthermal velocity is located around the top of the flaring loop, although \ion{Fe}{24} of 192.03 \AA\ is saturated.
Fortunately, we also have another \ion{Fe}{24} of 255.11 \AA\ data, which is not saturated.
By using \ion{Fe}{24} 255.11 \AA\ data, we obtain the same result that the maximum nonthermal velocity is located around the looptop of flare.
It is also remarkable that the blueshift component, $(x,y)=(-650,100)$, can be observed far from the location of nonthermal broadening.

Finally, we study the relationship of the location among the nonthermal broadening, the flare loop, and the flare ribbon.
We also discuss the temporal evolution of the relationship.
The temporal evolution of the relationship among the flare loop, the flare ribbon, and the nonthermal broadening is shown in Figure 11.
The color image, the green contour, and the black contour show AIA 131 \AA\ intensity, the nonthermal velocity of EIS 192.03 \AA, and AIA 1700 \AA\ intensity, respectively. 
The contour level of the nonthermal velocity is flexible.
In each panel of Figure 11(i-iv), the EIS raster scanning starts from 04:37:20, 04:43:10, 04:54:49, and 05:00:38, respectively.
During the thermal rich flare, the nonthermal broadening (green contour) are always observed around the looptop (or above-the-looptop).
The nonthermal broadening is seemed to be located apart from the flare ribbons.
The location of nonthermal broadening gradually moved to the higher location associated with the growth of the flare loop.

\subsection{Comparison with Intermediate and Thermal Rich Flare}

The intermediate flare observed on 23 September 2011 shows the typical characteristics of impulsive flare.
The lightcurves of GOES and high-energy electrons show the sharp peak.
The full-FOV nonthermal velocity of \ion{Fe}{24} peaked before the peak of the full-FOV intensity of \ion{Fe}{24} and GOES.
The location of nonthermal broadening is cospatial with the flare ribbons at the beginning of the flare.
After the microwave peak, the location of nonthermal broadening gradually moved toward the looptop.

The thermal rich flare observed on 25 September 2011 shows the typical characteristics of long duration flares.
The lightcurves of GOES and high-energy electrons show the gradual rising phase.
The full-FOV nonthermal velocity of \ion{Fe}{24} simultaneously peaked with the full-FOV intensity of \ion{Fe}{24} and GOES.
At the beginning of the flare, the location of nonthermal broadening is the looptop or above-the-looptop far from the flare ribbons.
During the flare, the nonthermal broadening is always located at the looptop or above-the-looptop.

The main difference between the intermediate and thermal rich flare is the location of the nonthermal broadening at the beginning of the flare.
The nonthermal broadening starts from the footpoints of the flare loop in the case of the intermediate flare, although the broadening starts at the lootop of the flare loop in the case of the thermal rich flare.
The relationship of timing between the full-FOV nonthermal broadening and the full-FOV intensity of \ion{Fe}{24} is also different between the intermediate and thermal rich flares.
The full-FOV nonthermal velocity peaked before the peak of the full-FOV intensity in the case of the intermediate flare, although the full-FOV nonthermal velocity and the full-FOV intensity peaked simultaneously.

\section{Other Flares and Microwave Source}
We investigate the same analysis for more three event to ensure the results which is discussed in the previous section.
We carry out full-FOV analysis in the same way to clarify the temporal evolution of high temperature plasma in the entire flare region by using \ion{Fe}{24} 192.03 \AA . 
In the event (a) which can be classified as intermediate flare, the full-FOV nonthermal velocity peaked before the peak of the full-FOV intensity in the case of the intermediate flare (not shown here).
On the other hand, in the event (c) which can be classified as thermal rich flare, the full-FOV nonthermal velocity and the full-FOV intensity peaked simultaneously (not shown here).
Therefore, the relationship of timing between the full-FOV nonthermal broadening and the full-FOV intensity of \ion{Fe}{24} shows the same trend in the other events. 
In the event (e) which can be classified as thermal poor flare, the duration of the flare is too short to discuss the relationship of timing between the intensity and nonthermal broadening from EIS observation, because the temporal resolution of EIS is long.  

We study the relationship of the location among the high-energy electrons, the nonthermal broadening, the flare loop, and the flare ribbon.
Figure 12 shows the intensity of AIA 131 \AA\ (color image), the nonthermal velcoity of EIS 192.03 \AA\ (green contour), the intensity of AIA 1700 \AA\ (black contour), and NoRH 17 GHz brightness temperature (red contour) at the peak of the full-FOV nonthermal velocity. 
The green contour represents 100, 140, 180 km sec$^{-1}$, and the red contour represents 75, 85, 95 \% of the maximum brightness temperature.
The AIA 131 \AA\ image correspond to the 10$^7$ K flare loop, and the AIA 1700 \AA\ contour correspond to the flare ribbons which is the footpoint of the flare loop.
It seems that, in the events (a) and (b) which can be classified as intermediate flares, the nonthermal broadening occurred at the footpoint.
The microwave source is located at the top of the flaring loop.
On the other hand, in the events (c) and (d)  which can be classified as the thermal rich flares, it seems that the nonthermal line broadening occurred at the looptop. 
The source of the microwave is located at the footpoint of the flare loop.
Further, in the event (e) which can be classified as the thermal poor flare, it seems that the nonthermal line broadening occurred at the footpoint. 
The microwave source is located at the top of the flaring loop.
The trend of the temporal evolution of the nonthermal line broadening was also the same discussed in the previous section (not shown here). 
All the flares, except for event (e) which has no data after the peak time of the microwave burst,  show that the nonthermal broadening is located at the top of the flaring loop after the microwave burst.

\section{Discussion}
\subsection{ Summary of the Results}
We analyzed the relationship between the location of EUV nonthermal broadening and high-energy particles during the large flares mainly by using the {\it Hinode}/EIS observation.
We also paid attention to the relationship between the location of nonthermal broadening and the thermal richness of the flares.
What we found are as follows:
(1) the nonthermal broadening of \ion{Fe}{24} occurred at the footpoint of the flare loop at the beginning of the intermediate flares,
(2) the source of the 17 GHz microwave is located at the top of the flaring loop at the beginning of the intermediate flares,
(3) the nonthermal broadening of \ion{Fe}{24} gradually moved toward the looptop after the microwave burst in the case of intermediate flares,
(4) the nonthermal broadening of \ion{Fe}{24} occurred at the top of the flaring loop at the beginning of the thermal rich flares,
(5) the source of the 17 GHz microwave is located at the footpoint of the flare loop at the beginning of the thermal rich flares,
(6) the nonthermal broadening of \ion{Fe}{24} is also located at the top of the flaring loop after the microwave burst in the case of thermal rich flares.
The results are summarized in Table 2

\subsection{ Nonthermal Broadening at Footpoints and Looptop Sources}
Since we have studied the location of the nonthermal broadening based on the imaging spectroscopic observation, we are in a better position to discuss the nature of 
the various sources of the nonthermal broadening.
As we mentioned in Section 1, the past observations indicated that the nonthermal broadening might occur at both the looptop and footpoint of the flare loop \citep[e.g.,][]{mariska1999}. 
Further, \cite{ranns2001}  found that the nonthermal broadening occurs after the HXR burst in the case of the impulsive flares.
On the other hand, they also found that the nonthermal broadening occurs before the HXR burst in the case of the long duration flares.
They concluded that the origin of the nonthermal broadening might be the evaporation flow at the footpoint of the flare loop in the case of impulsive flares,
although the origin of the nonthermal broadening might be the turbulence at the top of the flaring loop in the case of long duration flares.
We studied the location of the nonthermal broadening in both the case of impulsive and long duration flares.
Our results are certainly consistent with the past results which are from spatially unresolved observations.
Further, we found that some of the nonthermal broadening during the impulsive flares were clearly associated with the flare ribbons (Figure 6) and the blueshift component of \ion{Fe}{24} (Figure 5), which is the indication of the chromospheric evaporation flow.
Evaporation flows associated with nonthermal broadening are well observed at the footpoint of flare loops \citep[e.g.,][]{milligan2011}. 
The origin of nonthermal broadening is thought to be multiple flow speeds of evaporation, because generally its line profile is highly distorted \citep[e.g.,][]{imada2008}.
Therefore, it is plausible that the origin of the nonthermal broadening at the footpoint during the intermediate flare is the evaporation flow.
To discuss the origin of  the nonthermal broadening at the looptop in the case of the thermal rich flares, we study the detailed line profile of \ion{Fe}{24}.
Figure 13 shows the line profile integrated over 15 $\times$ 15 arcsec$^2$ around the peak position of the nonthermal velocity.
The + symbol represents the observation, and the solid line represents the Gaussian fitting result.
The line profile is almost symmetric (at least within FWHM), although there is an excess at the blue-wing. 
Further, it seems that the location of the nonthermal broadening is above-the-looptop rather than the top of the flaring loop, where we cannot observe any signature of evaporation flow (for example, flare ribbons, flows).
The nonthermal broadening at the looptop has a different characteristics from the nonthermal broadening at the footpoint.
Thus, it is plausible that the origin of the nonthermal broadening at the looptop might be the turbulence/wave which is  produced by the interaction between the reconnection outflow and the bright flare loop.

\subsection{ Thermal Rich vs. Thermal Poor Flares}
In our results, the thermal rich flares have a footpoint source of the microwave emission, and the other flares have a looptop source of the microwave emission.
This result is consistent with the discussions in \cite{kawate2011}.
They claimed that the thermal rich flares have microwave sources located near footpoints based on the microwave flux increasing from the center to limb.
On the other hand, the looptop microwave source may indicate that the high-energy electrons are trapped at the looptop either by wave scattering or by magnetic mirroring force.

We describe our interpretation of observational results in Figure 14.
The microwave sources are found at the looptop/footpoint of the flares in the case of intermediate/thermal rich flares, respectively.
Furthermore, we observe the nonthermal broadening at the footpoint/looptop in the case of intermediate/thermal rich flares, respectively.
We think that the nonthermal broadening may give a clue for the presence of turbulence/wave that plays an important role in the pitch angle scattering of the high energy electrons.
The pitch angle distribution of high-energy electron in the case of the thermal rich flares should be isotropic at the looptop.
Then, the most of high-energy, even low-energy, electrons can travel along the flare loop and gradually penetrate into the chromosphere,
because the loss cone is refilled by the continuous pitch angle scattering at the looptop during the bouncing motion.
On the other hand, in the case of intermediate and thermal poor flares, the only electron whose pitch angle is inside the loss cone can precipitate into the chromosphere at the beginning of the flares.
However, to prove our speculation, whether or not the turbulence can really interact with electrons to play a role in the 
electron pitch angle scattering should be known.  It is thus an important task in future studies to find out the spatial scale of the 
turbulence/waves underlying the nonthermal broadening presented in this study.

\acknowledgments 
We thank T. Watanabe, H. Hara, K. Ichimoto, V. Melnikov, and K. Shibata for fruitful discussions.

Hinode is a Japanese mission developed and launched by ISAS/JAXA, collaborating with NAOJ as a domestic 
partner, NASA and STFC (UK) as international partners. Scientific operation of the Hinode mission is conducted by 
the Hinode science team organized at ISAS/JAXA. This team mainly consists of scientists from institutes in the 
partner countries. Support for the post-launch operation is provided by JAXA and NAOJ (Japan), STFC (U.K.), 
NASA, ESA, and NSC (Norway).

\begin{table*}
\begin{center}
\caption{Flare Events. }
\begin{tabular}{crrrrrrrrrrr}
\tableline\tableline
event & microwave peak time & flare position & GOES class & TEI \\
\tableline
(a) & 2011/02/16 01:36:53 & (373,-241) & M1.0 & 0.06 \\ 
(b) & 2011/09/23 23:53:53 & (-859,162) & M1.9 & 0.05 \\ 
(c) & 2011/09/25 02:32:35 & (-707,147) & M4.4 & 0.73 \\ 
(d) & 2011/09/25 04:52:55 & (-653, 98) & M7.4 & 0.58 \\ 
(e) & 2012/07/14 04:54:48 & (319, -319) & M1.0 & -0.52 \\
\tableline
\end{tabular}
\end{center}
\label{t1}
\end{table*}

\begin{figure}
 \begin{center}
  \includegraphics[width=140mm]{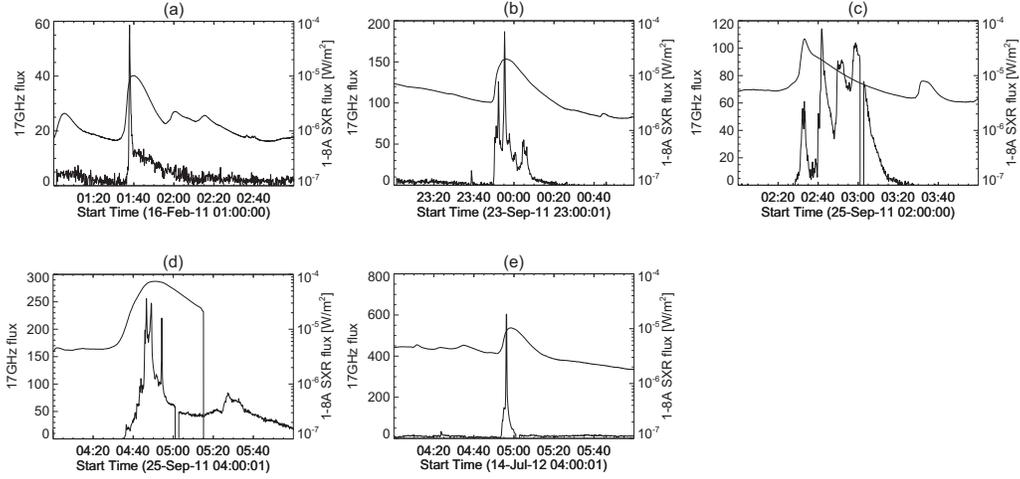}
 \end{center}
 \caption{Lightcurves of SXR (thin) and microwaves (thick) during the flare events.}
 \label{f2-1}
\end{figure}

\begin{figure}
 \begin{center}
  \includegraphics[width=100mm]{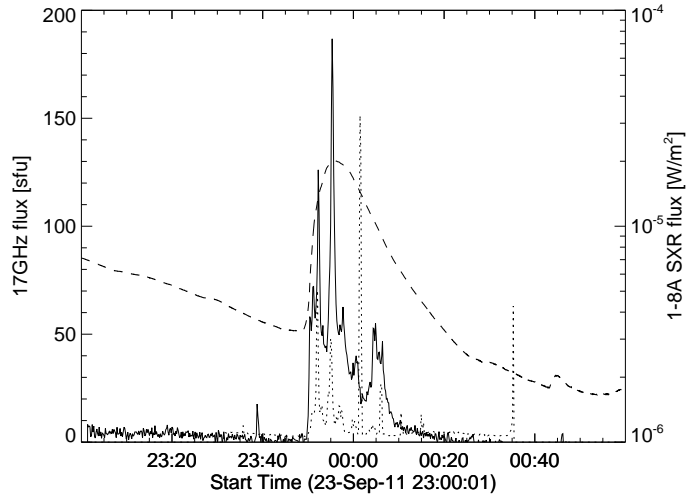}
 \end{center}
 \caption{Lightcurves of 17 GHz microwaves (solid), GOES (dashed), and HXR 25-50 keV (dotted) during the intermediate flare on 23rd September 2011 (event b).}
 \label{f2-3-1}
\end{figure}

\begin{figure}
 \begin{center}
  \includegraphics[width=100mm]{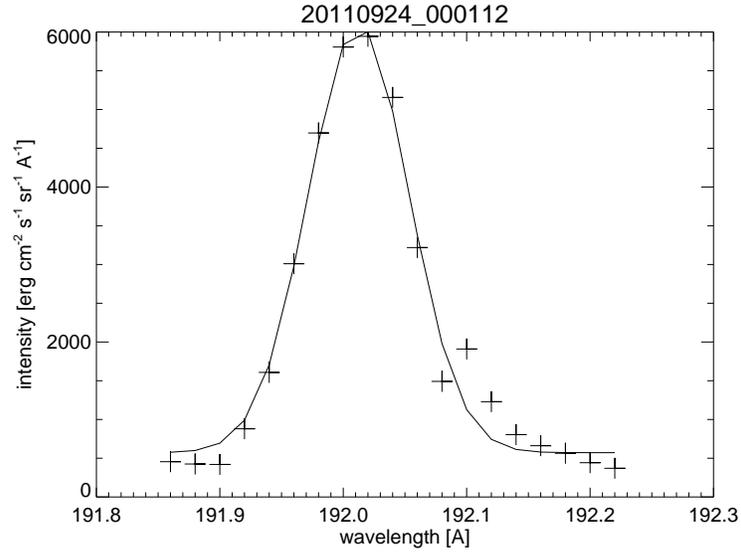}
 \end{center}
 \caption{Example of \ion{Fe}{24} line profile integrated over the EIS FOV  during the intermediate flare (+: observation, solid line: Gaussian fitting).}
 \label{f2-3-2}
\end{figure}

\begin{figure}
 \begin{center}
  \includegraphics[width=140mm]{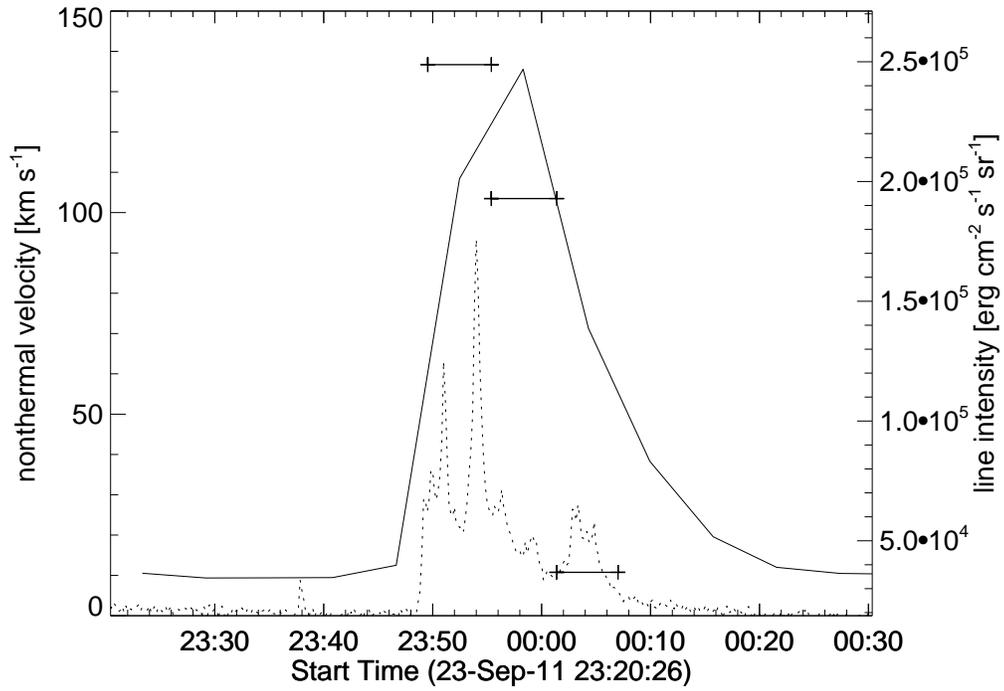}
 \end{center}
 \caption{Temporal evolution of the intermediate flare event (horizontal line: \ion{Fe}{24} full-FOV nonthermal velocity, solid line: \ion{Fe}{24} full-FOV intensity integrated from 191.90 to 192.15 \AA, dotted line: 17 GHz microwave).}
 \label{f2-3-3}
\end{figure}

\begin{figure}
 \begin{center}
  \includegraphics[width=140mm]{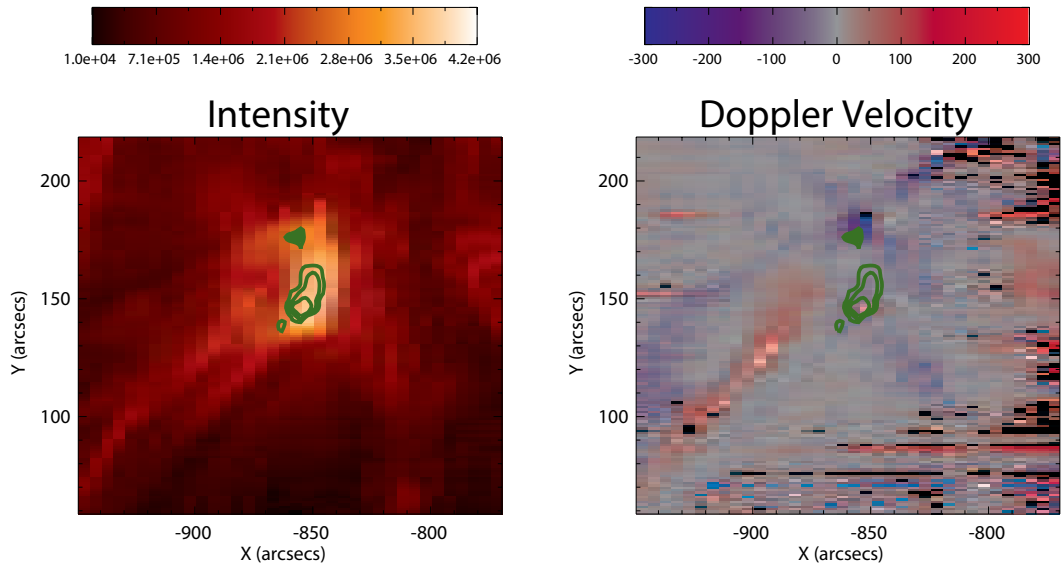}
 \end{center}
 \caption{\ion{Fe}{24} intensity (erg cm$^{-2}$ sec$^{-1}$ str$^{-1}$ \AA$^{-1}$), velocity (km sec$^{-1}$), and non thermal velocity (km sec$^{-1}$) at the peak time of nonthermal velocity (see Figure 4) during the  intermediate flare. 
Green contours represent the nonthermal velocity of 100, 140, 180 km sec$^{-1}$.}
 \label{f2-3-4}
\end{figure}

\begin{figure}
 \begin{center}
  \includegraphics[width=140mm]{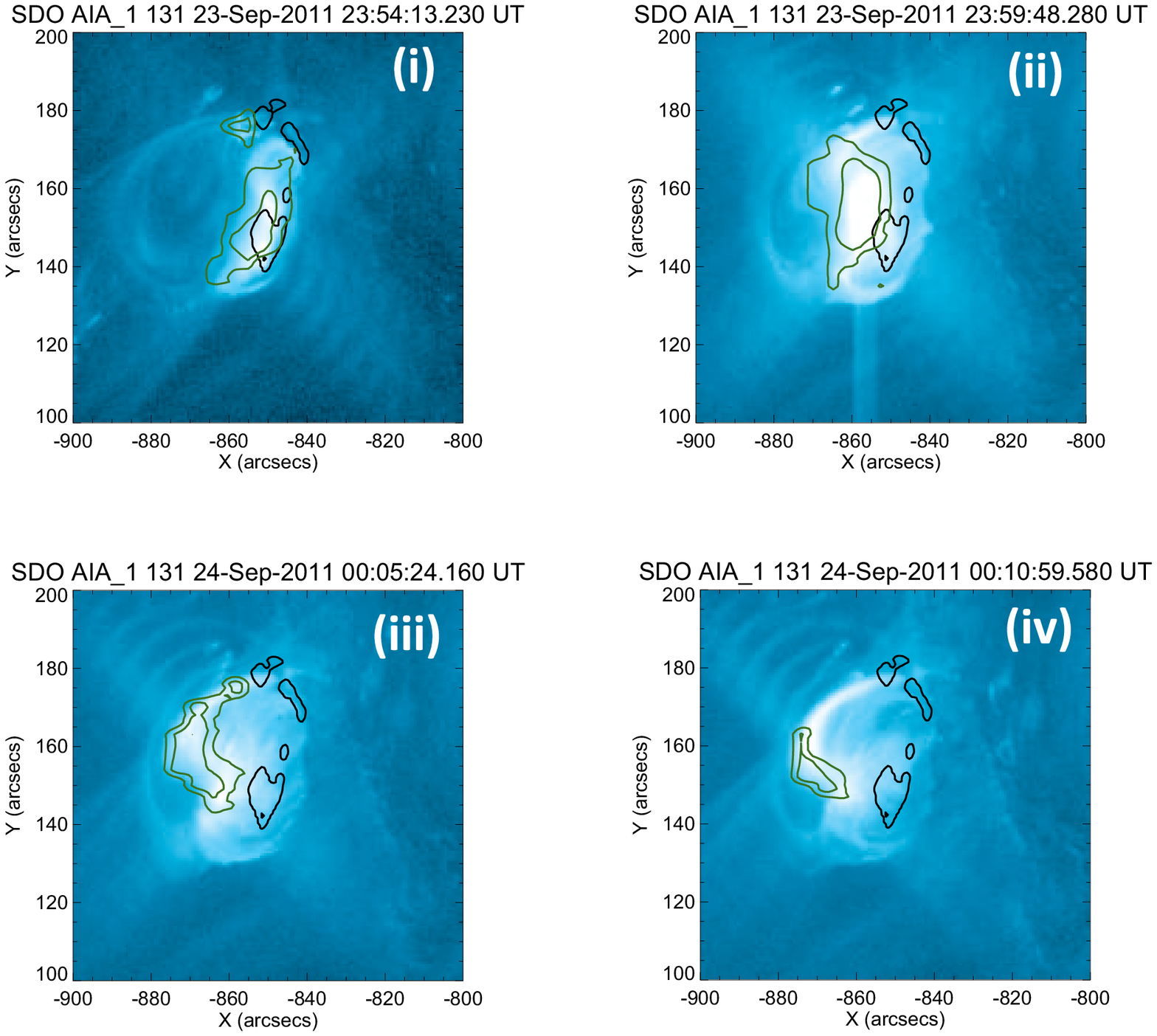}
 \end{center}
 \caption{Temporal evolution of the nonthermal line broadening during the intermediate flare. The color image, the green contour, and the black contour are the AIA 131 (\AA) intensity map, the nonthermal velocity of \ion{Fe}{24} (192.03 \AA), and the AIA 1700 (\AA) intensity, respectively.}
 \label{f2-3-5}
\end{figure}

\begin{figure}
 \begin{center}
  \includegraphics[width=140mm]{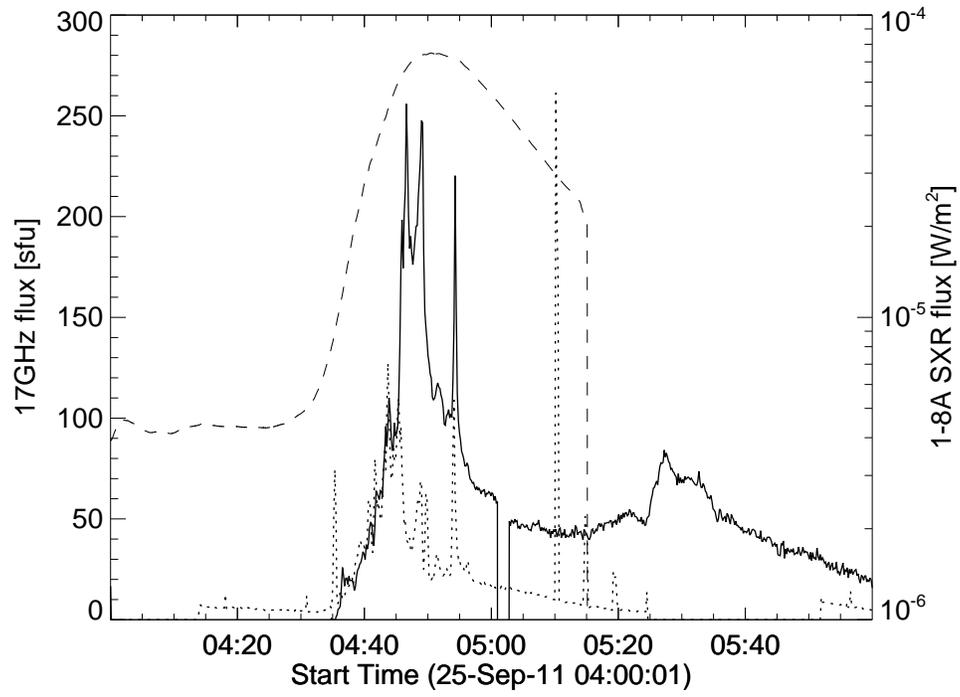}
 \end{center}
 \caption{Lightcurves of 17 GHz microwaves (solid), GOES (dashed), and HXR 25-50 keV (dotted) during the thermal rich flare on 25th September 2011 (event d).}
 \label{f2-4-1}
\end{figure}

\begin{figure}
 \begin{center}
  \includegraphics[width=140mm]{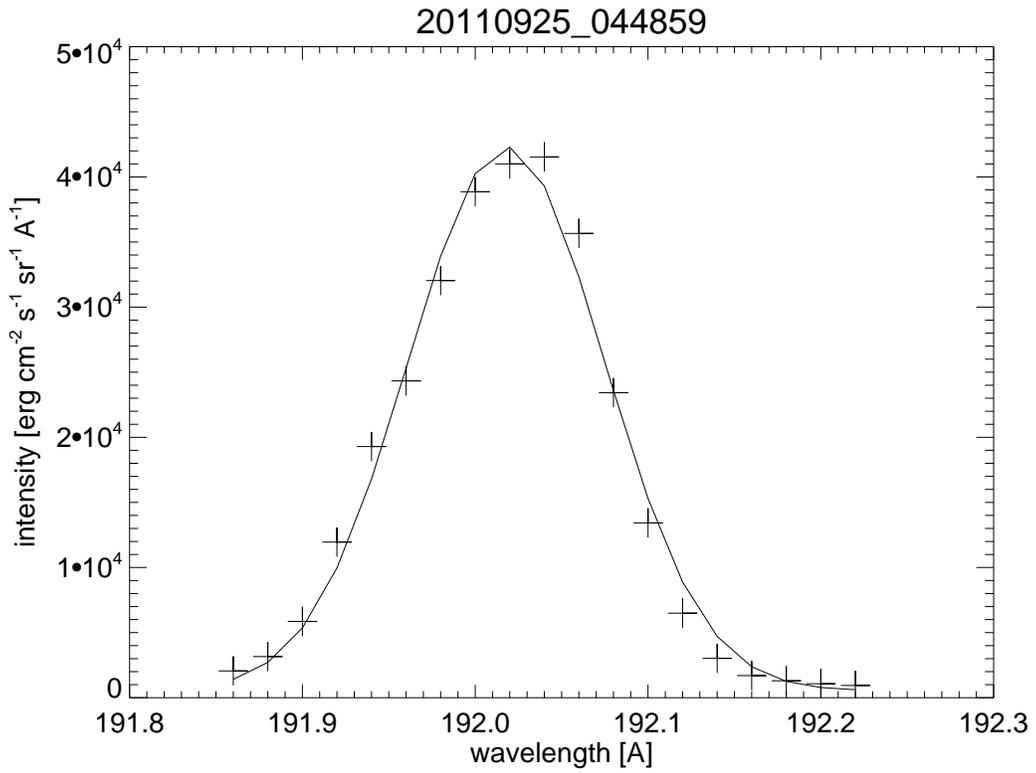}
 \end{center}
 \caption{Example of \ion{Fe}{24} line profile integrated over the EIS FOV during the thermal rich flare (+: observation, solid line: Gaussian fitting).}
 \label{f2-4-2}
\end{figure}

\begin{figure}
 \begin{center}
  \includegraphics[width=140mm]{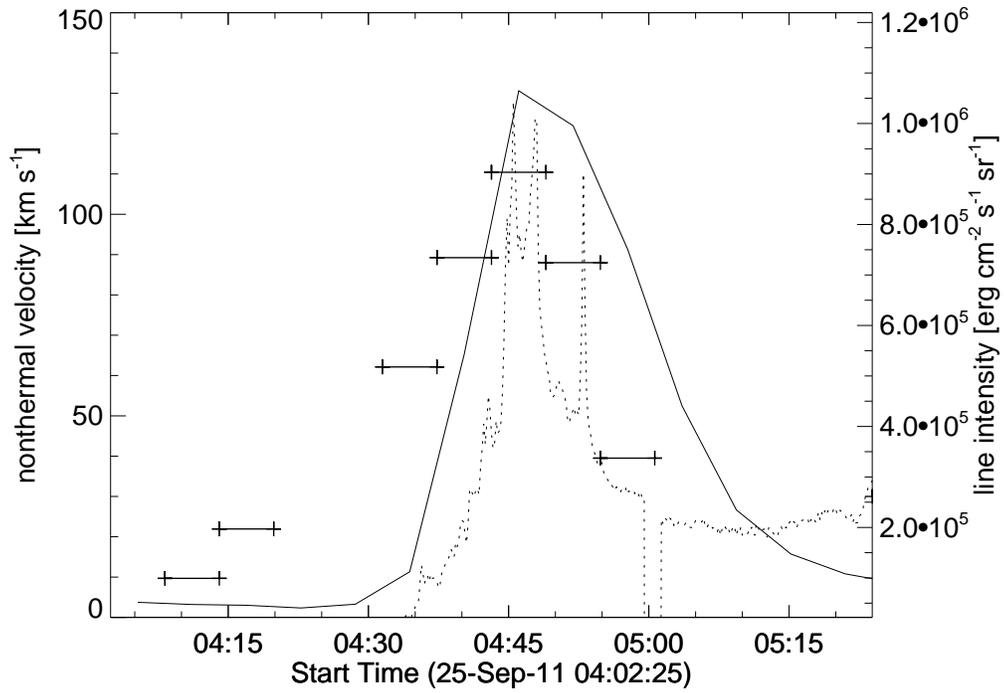}
 \end{center}
 \caption{Temporal evolution of the thermal rich flare event (horizontal line: \ion{Fe}{24} full-FOV nonthermal velocity estimated from integrated line profile over EIS FOV, solid line: \ion{Fe}{24} full-FOV intensity integrated from 191.90 to 192.15 \AA, dotted line: 17 GHz microwave).}
 \label{f2-4-3}
\end{figure}

\begin{figure}
 \begin{center}
  \includegraphics[width=140mm]{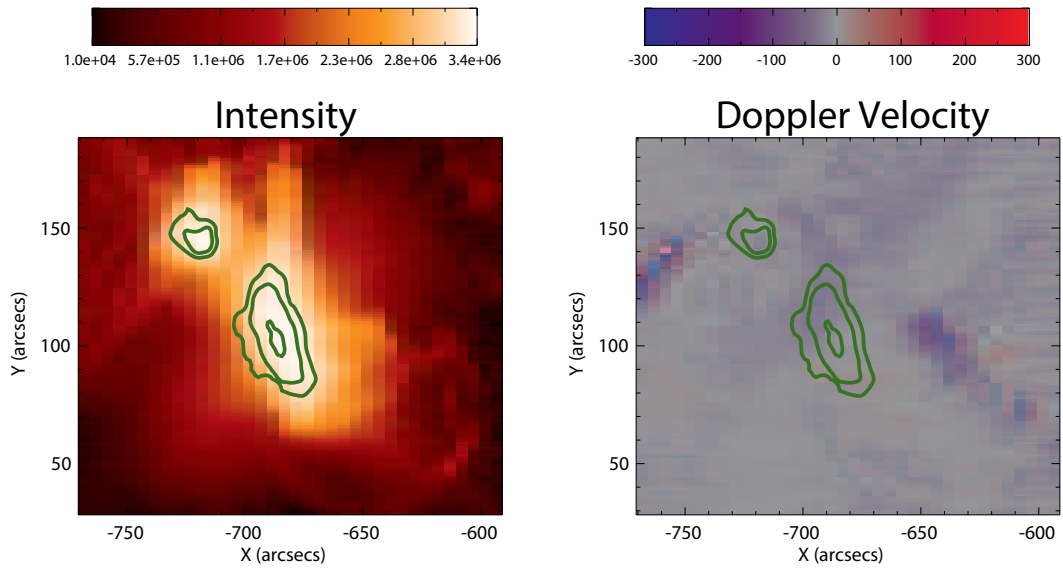}
 \end{center}
 \caption{\ion{Fe}{24} intensity (erg cm$^{-2}$ sec$^{-1}$ str$^{-1}$ \AA$^{-1}$), velocity (km sec$^{-1}$), and nonthermal velocity (km sec$^{-1}$) at the peak time of nonthermal velocity (see Figure 9) during the thermal rich flare. Green contour represents the nonthermal velocity of 100, 140, 180 km sec$^{-1}$.}
 \label{f2-4-4}
\end{figure}

\begin{figure}
 \begin{center}
  \includegraphics[width=140mm]{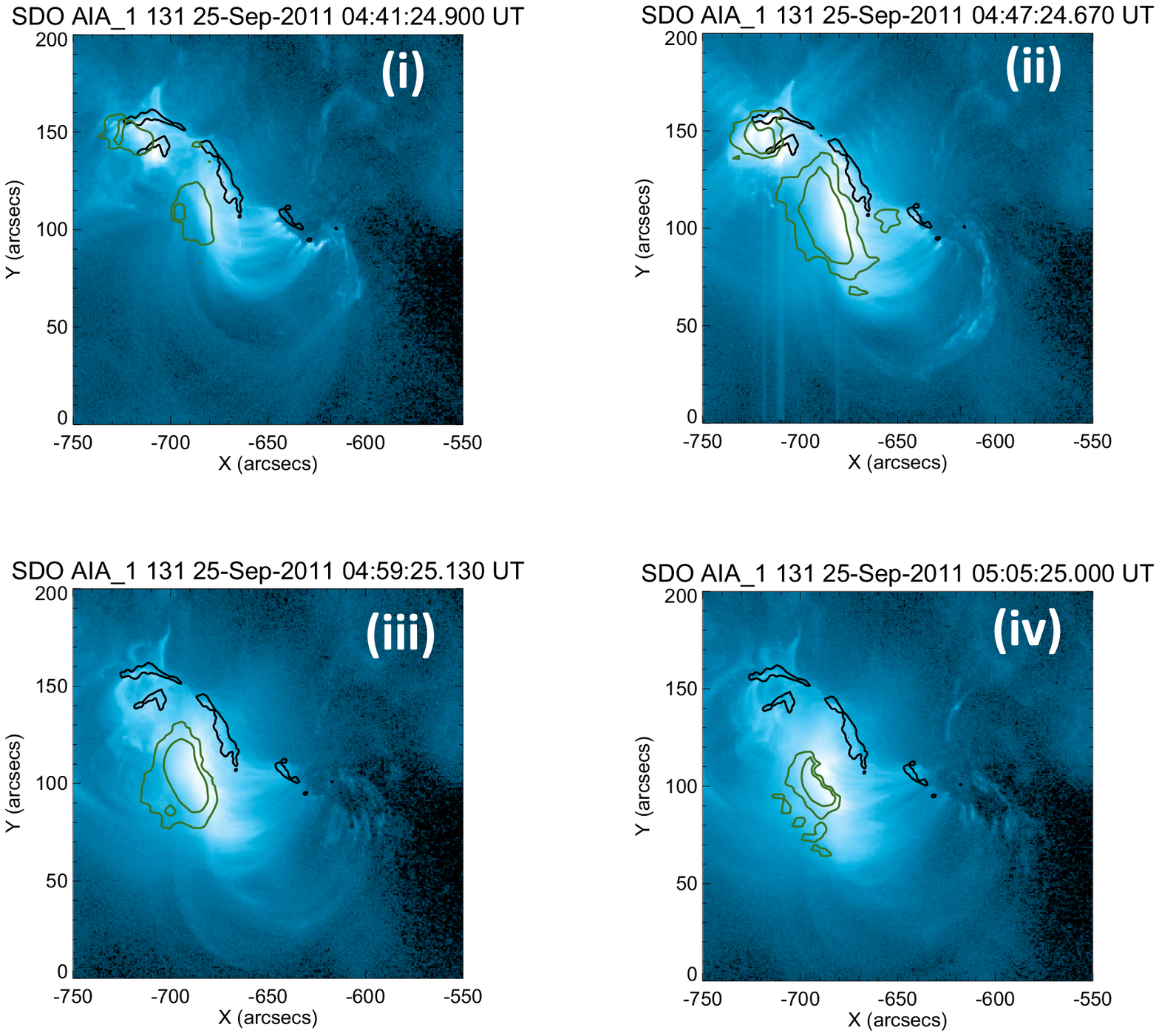}
 \end{center}
 \caption{Temporal evolution of the nonthermal line broadening during the thermal rich flare. The color image, the green contour, and the black contour are the AIA 131 (\AA) intensity map, the nonthermal velocity of \ion{Fe}{24} (192.03 \AA), and the AIA 1700 (\AA) intensity, respectively.}
 \label{f2-4-5}
\end{figure}

\begin{figure}
 \begin{center}
  \includegraphics[width=100mm]{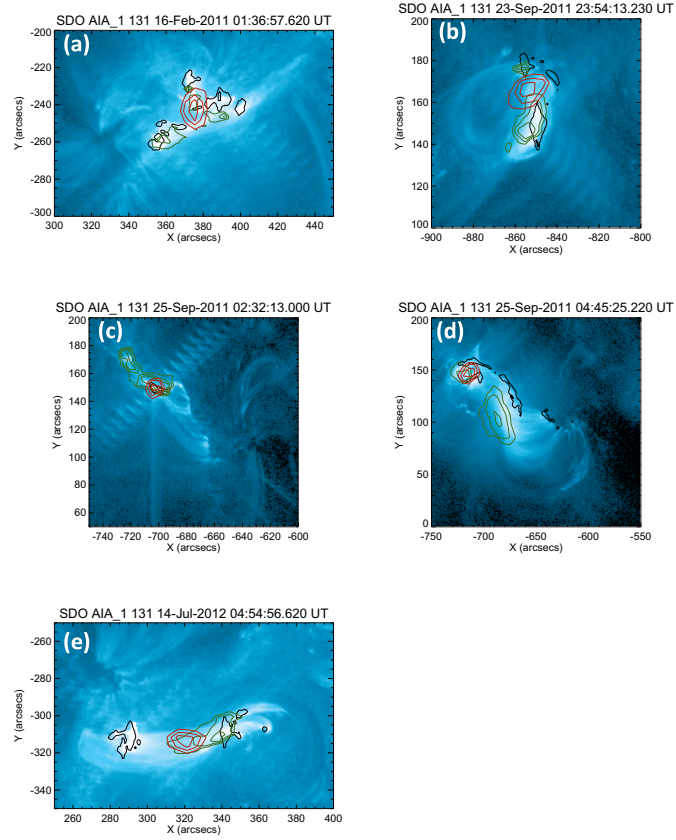}
 \end{center}
 \caption{Relationship among the location of flare loop (AIA 131 \AA : image), nonthermal line broadening (EIS \ion{Fe}{24}: green contour), microwave (NoRH 17 GHz: red contour), and flare ribbon (AIA 1700 \AA: black contour) at the nonthermal peak of the flares.}
 \label{f3}
\end{figure}

\begin{figure}[htbp]
 \begin{center}
  \includegraphics[width=140mm]{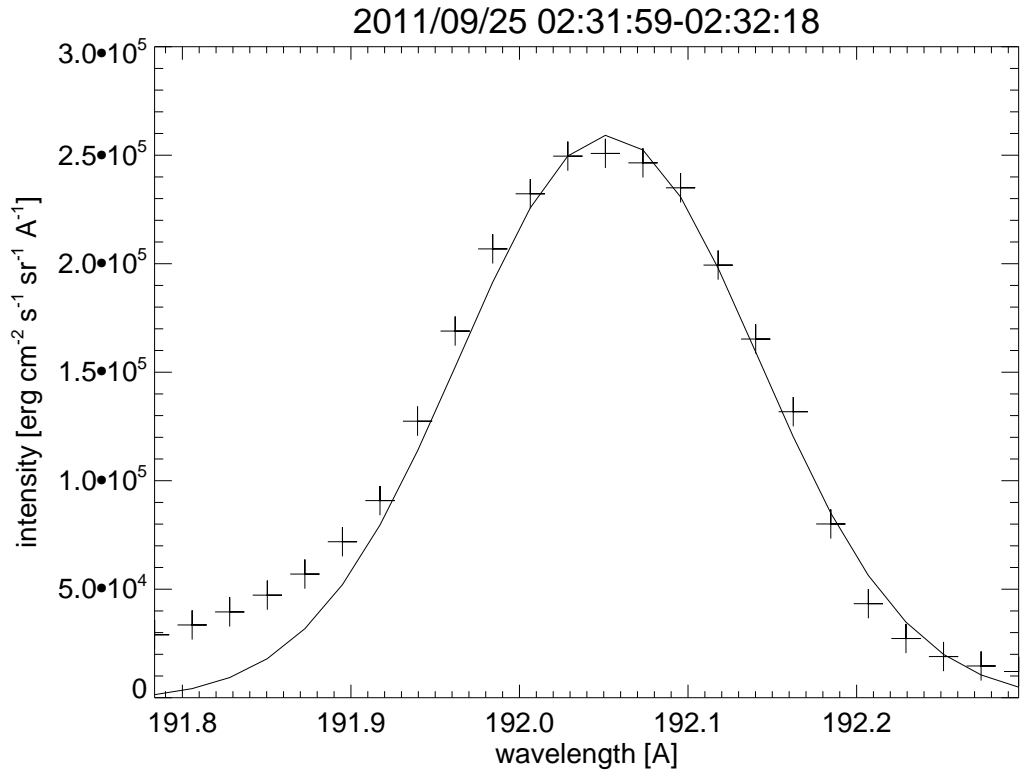}
 \end{center}
 \caption{\ion{Fe}{24} line spectra integrated 15 $\times$ 15 arcsec$^2$ around the peak position of the nonthermal broadening at the beginning of the thermal rich flare (event (c)) (+: observation, solid line: Gaussian fitting).}
 \label{f4-1}
\end{figure}

\begin{figure}[htbp]
 \begin{center}
  \includegraphics[width=140mm]{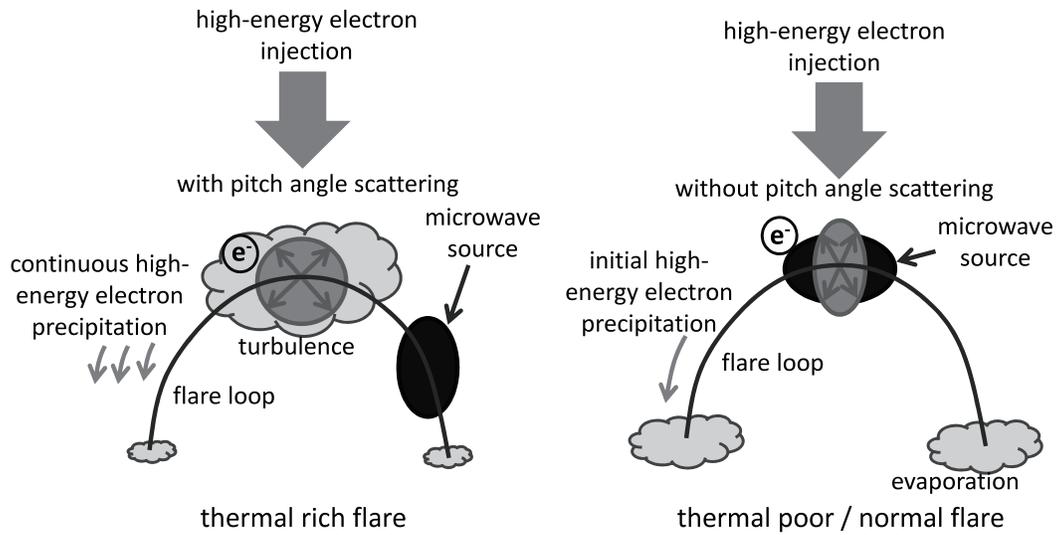}
 \end{center}
 \caption{Schematic illustration of the difference between thermal rich flares and the other flares. The gray arrow in a circle means the pitch angle distribution of high-energy electron.}
 \label{f4-2}
\end{figure}

\begin{table*}
\begin{center}
\caption{Summary of Observations. }
\begin{tabular}{crrrrrrrrrrr}
\tableline\tableline
event & GOES class & TEI & impulsive & nonthermal broadening& 17 GHz source \\
\tableline
(a) &  M1.0 & 0.06 & yes & footpoint & looptop\\
(b) & M1.9 & 0.05 & yes & footpoint & looptop\\ 
(c) & M4.4 & 0.73 & no & looptop & footpoint\\
(d) & M7.4 & 0.58 & no & looptop & footpoint\\
(e) & M1.0 & -0.52 & yes & footpoint & looptop\\
\tableline
\end{tabular}
\end{center}
\label{t2}
\end{table*}



\begin{thebibliography}{}
\bibitem[Ambrosiano et al.(1988)]{amb} Ambrosiano, J., Mattaeus, W. H., Goldsteins, M. L. \& Plante, D. 1988, \jgr, 93, 14,383
\bibitem[Blandford  and Ostriker (1978)]{bla} Blandford, L. \& Ostriker, J. P. 1978, \apjl, 221, L29
\bibitem[Carmichael (1964)]{car} Carmichael, H. 1964, in NASA Special Publication 50, The Physics of Solar Flares, ed. W. N. Hess (Washington, DC: NASA), 451
\bibitem[Culhane et al. (1991)]{culhane1991}Culhane, J. L., Hiei, E., Doscheck, G. A. et al. 1991, \solphys, 136, 89
\bibitem[Culhane et al. (2007)]{culhane2007}Culhane, J. L., Harra, J. K., James, A. M. et al. 2007, \solphys, 243, 19
\bibitem[Doschek et al. (2007)]{dos}Doschek, G. A., Mariska, J. T., Warren, H. P., et al.  2007, \apjl, 667, 109
\bibitem[Hara et al. (2008)]{hara2008}Hara, H., Watanabe, T., Matsuzaki, K. et al. 2008, \pasj, 60,275
\bibitem[Harra-Murnion et al. (1997)]{harra1997}Harra-Murnion, L. K., Akita, K. \& Watanabe, T. 1997, \apjl, 479, 464
\bibitem[Hirayama (1974)]{hir} Hirayama, T. 1974, \solphys, 34, 323
\bibitem[Imada et al. (2007)]{ima} Imada, S., Nakamura, R., Daly, P. W., et al. 2007, \jgr, 112, A03202
\bibitem[Imada et al. (2008)]{imada2008}Imada, S., Hara, H., Watanabe, T. et al. 2008, \apjl, 679, 155
\bibitem[Imada et al. (2009)]{ima3}Imada, S., Hara, H. \& Watanabe, T.  2009, \apjl, 705, L208
\bibitem[Imada et al. (2011)]{ima2} Imada, S., Hirai, M., Hoshino, M. \& Mukai, T. 2011, \jgr, 116, A08217
\bibitem[Kawate et al. (2011)]{kawate2011}Kawate, T., Asai, A. \& Ichimoto, K. 2011, \pasj, 63, 1251
\bibitem[Kamio et al.(2010)]{kam}Kamio, S., Hara, H., Watanabe, T., Fredvik, T. \& Hansteen, V. 2010, \solphys, 266, 209 
\bibitem[Kliem (1994)]{kli}Kliem, B. 1994, \apjs, 90, 719
\bibitem[Kopp \& Pneuman (1976)]{kop} Kopp, R.A. \& Pneuman, G.W. 1976, \solphys, 50, 85
\bibitem[Kosugi et al. (1991)]{kosugi1991} Kosugi, T., Makishima, K., Murakami, T., et al. 1991, \solphys, 136, 17
\bibitem[Kosugi et al. (2007)]{kosugi2007} Kosugi, T., Matsuzaki, K., Sakao, T. et al. 2007, \solphys, 243, 3
\bibitem[Lemen et al. (2012)]{lemen2012}Lemen, J. R., Title, A. M., Akin, D. J. et al. 2012, \solphys, 275, 17
\bibitem[Lin et al. (2002)]{lin2002}Lin, R. P., Dennis, B. R., Hurford, G. J. et al. 2002, \solphys, 210, 3
\bibitem[Masuda et al. (1994)]{mas} Masuda, S., Kosugi, T., Hara, H., Tsuneta, S. \& Ogawara, Y. 1994, \nat, 371, 495
\bibitem[Mariska et al. (1993)]{mariska1993}Mariska, J. T., Doschek, G. A. \& Bentley, R. D. 1993, \apj, 419, 418
\bibitem[Mariska \& McTiernan (1999)]{mariska1999}Mariska, J. T. \& McTiernan, J. M. 1999, \apj, 514, 484
\bibitem[Melrose (1974)]{mel} Melrose, D. B. 1974, \solphys, 116, A08217
\bibitem[Milligan (2011)]{milligan2011}Milligan, R. O. 2011, \apj, 740, 70
\bibitem[Minoshima et al. (2010)]{min}Minoshima, T., Masuda, S., \& Miyoshi, Y. 2010, \apj, 714, 332
\bibitem[Nakajima et al. (1985)]{nakajima1985}Nakajima, H., Sekiguchi, H., Sawa, M., Kai, K. \& Kawashima, S. 1985, \pasj, 37, 163
\bibitem[Nakajima et al. (1994)]{nakajima1994}Nakajima, H., Nishio, M., Enome, S., et al. 1994, Proc. IEEE, 82, 705 
\bibitem[ {\O}ieroset et al. (2002)]{oie} {\O}ieroset, M., Lin, R. P., Phan, T. D., Larson, D. E. \& Bale, S. D. 2002, \prl, 89, 195001
\bibitem[Pneuman et al.(1981)]{pne} Pneuman,~G.W. 1981, In {\it Solar Flare Magnetohydrodynamics}, ed. ER Priest,  379, London:Gordon \& Breach
\bibitem[Ranns et al. (2000)]{ranns2000}Ranns, N. D. R., Matthews, S. A., Harra, L. K. \& Culhane, J. L. 2000, \aap, 364, 859
\bibitem[Ranns et al. (2001)]{ranns2001}Ranns, N. D. R., Harra, L. K., Matthews, S. A., \& Culhane, J. L. 2001, \aap, 379, 616
\bibitem[Rapley et al. (1977)]{rapley1977}Rapley, C. G., Culhane, J. L., Acton, L. W. et al., 1977, Review of Scientific Instruments, 48, 1123
\bibitem[Sarris et al. (1976)]{sar}Sarris, E. T., Krimigis, S. M., Bostrom, C. O., Iijima, T. \& Armstrong, T. P. 1976, \grl, 3, 437
\bibitem[Sturrock (1966)]{stu} Sturrock, P.A. 1966, \nat, 211, 695
\bibitem[Torii et al. (1979)]{torii1979}Nakajima, H., Nishio, M., Enome, S., et al. 1994, Proc. IEEE, 82, 705
\bibitem[Tsuneta \& Naito (1998)]{tsu} Tsuneta, S. \& Naito, T. 1998, \apjl, 495, 67
\bibitem[Watanabe (2008)]{watanabe2008}Watanabe, T. 2008, ASP Conference Series, 397, 139
\bibitem[Watanabe et al. (2012)]{wat} Watanabe, K.,  Masuda, S. \& Segawa, T., \solphys, 279, 317, 2012
\end{thebibliography}
\end{document}